\documentclass[a4paper,10pt,twocolumn]{article}

\usepackage[english]{babel}
\usepackage[utf8]{inputenc}
\usepackage[T1]{fontenc}

\usepackage[top=1.5cm, left=1.5cm, right=1.5cm, bottom=1.5cm]{geometry}

\renewenvironment{abstract}{\bf\small {\em\ Abstract---}}{}

\usepackage{amsfonts,amssymb,amsmath,amsthm}
\usepackage{subfigure}
\usepackage{graphicx}
\usepackage[footnotesize]{caption}
\usepackage{cite}

\usepackage{xcolor,url}
\usepackage{algorithm,algorithmic} 
\newcommand{\normF}[1]{\left|\left| #1 \right|\right|_\mathcal{F}}

\newcommand{\R}{\mathbb{R}}
\newcommand{\N}{\mathbb{N}}

\newcommand{\X}{\mathbb{X}}
\newcommand{\MPMP}[1]{\textcolor{black}{#1}}

\usepackage{bbold}

\title{Faster-than-fast NMF using random projections and Nesterov iterations}

\author{Farouk Yahaya, Matthieu Puigt, Gilles Delmaire, and Gilles Roussel\\
  \footnotesize Univ. Littoral C\^ote d'Opale, LISIC -- EA 4491, F-62228 Calais, France} \date{\empty} 

\begin{document}

\maketitle

\begin{abstract}%
Random projections have been recently implemented in Nonnegative Matrix Factorization (NMF) to speed-up the NMF computations, with a negligible loss of performance. In this paper, we investigate the effects of such projections when the NMF technique uses the fast Nesterov gradient descent (NeNMF). We experimentally show the randomized subspace iteration to significantly speed-up NeNMF.
\end{abstract}

\section{Introduction}\label{sec:introduction}

\MPMP{Modern latent variable analysis methods---e.g., sparse approximation, robust principal component analysis, dictionary learning---have been massively investigated for more than two decades and were successfully applied to signal, image, or video processing, and to machine learning. Among these techniques, Nonnegative Matrix Factorization (NMF) attracted a lot of interest from the scientific community since the pioneering work in \cite{Paa94,Lee_1999}. Indeed, it usually provides more interpretable results than methods without any sign constraint (e.g., independent component analysis) \cite{Gillis_2014} and it was successfully applied to many fields, e.g., audio signals \cite{Fevotte_2018}, hyperspectral unmixing \cite{Bioucas_2012}, or environmental data processing \cite{Puigt_2017}. NMF consists of estimating two $n \times p$ and $p \times m$ nonnegative matrices $G$ and $F$, respectively, from a $n\times m$ nonnegative matrix $X$ such that \cite{Wang_2013}
\begin{equation}
X \simeq G \cdot F.\label{eq:1}
\end{equation}
NMF usually consists of solving alternating subproblems, i.e.,
\begin{align}
\hat{G} & =  \arg\min_{G \geq 0} \normF{X - G \cdot F}, \label{eq:NMF_update1} \\
\hat{F} & =  \arg\min_{F \geq 0} \normF{X - G \cdot F}, \label{eq:NMF_update2}
\end{align}
using, e.g., Multiplicative Updates (MU)
\cite{Lee_1999}, Hierarchical Alternating Least Squares (HALS) \cite{Cichocki_2007}, Alternating Nonnegative Least Squares (ANLS)
\MPMP{\cite{Kim_2008}}, or Projected Gradient (PG)
\cite{Lin_2007}. Additionally, some authors incorporated some extra-information in the NMF model \cite{Wang_2013}, e.g., weights \cite{Ho_2008,Guillamet_2003}, 
sparsity assumptions \cite{Hoyer_2004,Dorffer_2018}, sum-to-one constraints \cite{Lanteri_2010}, specific matrix structures \cite{Meganem_2014,Dorffer_2016b}, or information \cite{Lin_2007,Choo_2014,Limem_2014a,Plouvin_2014,Limem_2014b,Dorffer_2018}. With the Big Data era, computational time reduction of NMF is particularly investigated, e.g., through optimal solvers \MPMP{\cite{Guan_2012}}, distributed strategies \cite{Liu_2010}, online estimation \cite{Mairal_2010}, or randomization \cite{Wang_2010,Tepper_2016,Erichson_2018}.}
The latter consists of reducing the size of some matrices through random projections---see, e.g., \cite{Halko_2011} for a comprehensive review---and thus to speed up the NMF computations\footnote{The authors in \cite{Tepper_2016,Erichson_2018} also proposed randomized techniques for separable NMF, which is out of the scope of this paper.}. However, the methods in \cite{Wang_2010,Tepper_2016,Erichson_2018} are based on MU, PG, or HALS. In this paper, we aim to investigate the benefits of compressing the data when a fast NMF solver using Nesterov iterations (NeNMF) \cite{Guan_2012} is used. Indeed, this approach was found to be among the fastest techniques in \cite{Sobral_2015}\footnote{See for example the CPU-time consumptions of \cite{Sobral_2015} at \url{https://github.com/andrewssobral/lrslibrary}.}. 

\MPMP{The remainder of the paper reads as follows. We recall the principles of NMF with Nesterov gradient descent in Sect.~\ref{sec:second-section}. Section~\ref{sec:first-section} introduces our proposed compressed NeNMF method\footnote{The Matlab code used in this paper is available at \url{https://gogs.univ-littoral.fr/puigt/Faster-than-fast_NMF}.} whose performance is investigated in Sect.~\ref{sec:third-section}. Lastly, we conclude and discuss about future directions in Sect.~\ref{sec:conclusion}.}

\section{NMF with Nesterov iterations}
\label{sec:second-section}
We firstly briefly recall the principles of the NeNMF method using Nesterov optimal gradient \cite{Guan_2012}. As 
\MPMP{explained above}, for a fixed $n \times m$ nonnegative data matrix $X$, NMF consist\MPMP{s} of finding \MPMP{both} the $n \times p$ and $p  \times m$ matrices $G$ and $F$ \MPMP{which} provide the best low-rank approximation of $X$ (\ref{eq:1}). 
NeNMF \cite{Guan_2012} iteratively and alternately solves (\ref{eq:NMF_update1}) and (\ref{eq:NMF_update2}) by applying in an inner loop the Nesterov accelerated gradient descent \cite{Nesterov_1983}. To update a factor, say $F$, the latter initializes $Y_0 \triangleq F^t$---where $t$ is an NeNMF outer iteration index---and considers a series $\alpha_k$ defined as $\alpha_0= 1$, and $\alpha_{k+1} = \frac{1 + \sqrt{4 \alpha_k^2 + 1}}{2}, \, \forall k \in \N.$ For each inner loop index $k$, the Nesterov gradient descent then computes an update $F_{k}$ of $F$ with a single gradient descent of $Y_k$, and then slides it in the direction of $F_{k-1}$---with weights from the series $\alpha_k$---to derive $Y_{k+1}$. Using the Karush-Kuhn-Tucker conditions, a stopping criterion---considering both a maximum number $\text{Max}_{\text{iter}}$ of iterations and a 
gradient bound---is proposed in \cite{Guan_2012}, thus yielding $F^{t+1} = Y_K$, where $Y_K$ is the last iterate of the above inner iterative gradient descent. The same strategy is applied to $G$. As shown in \cite{Guan_2012,Sobral_2015}, NeNMF is among the fastest state-of-the-art NMF techniques and is less sensitive to the matrix size than classical techniques, e.g., MU or PG.
\begin{figure*}
\centering
\includegraphics[width=.265\linewidth]{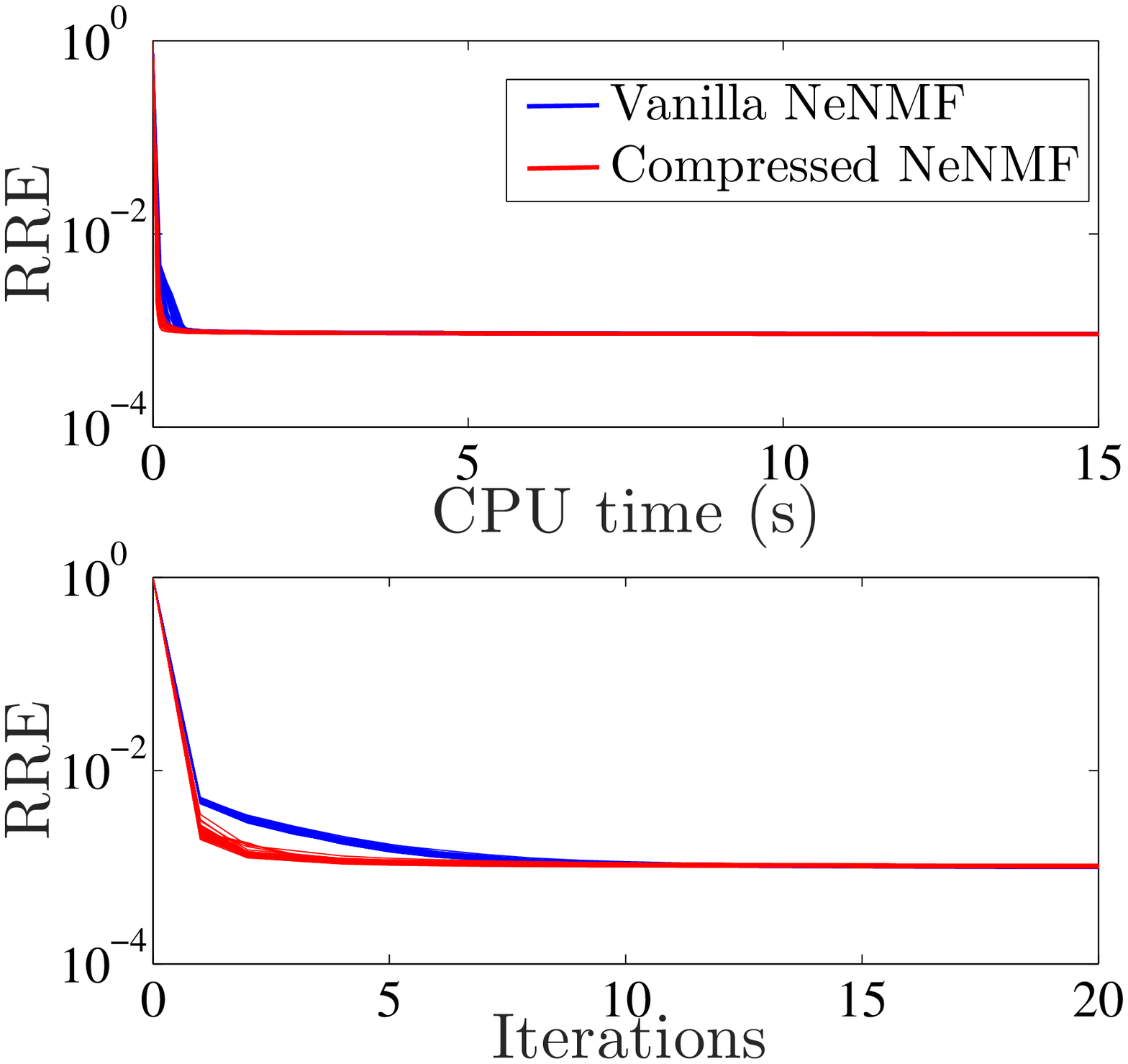}\hfill 
\includegraphics[width=.265\linewidth]{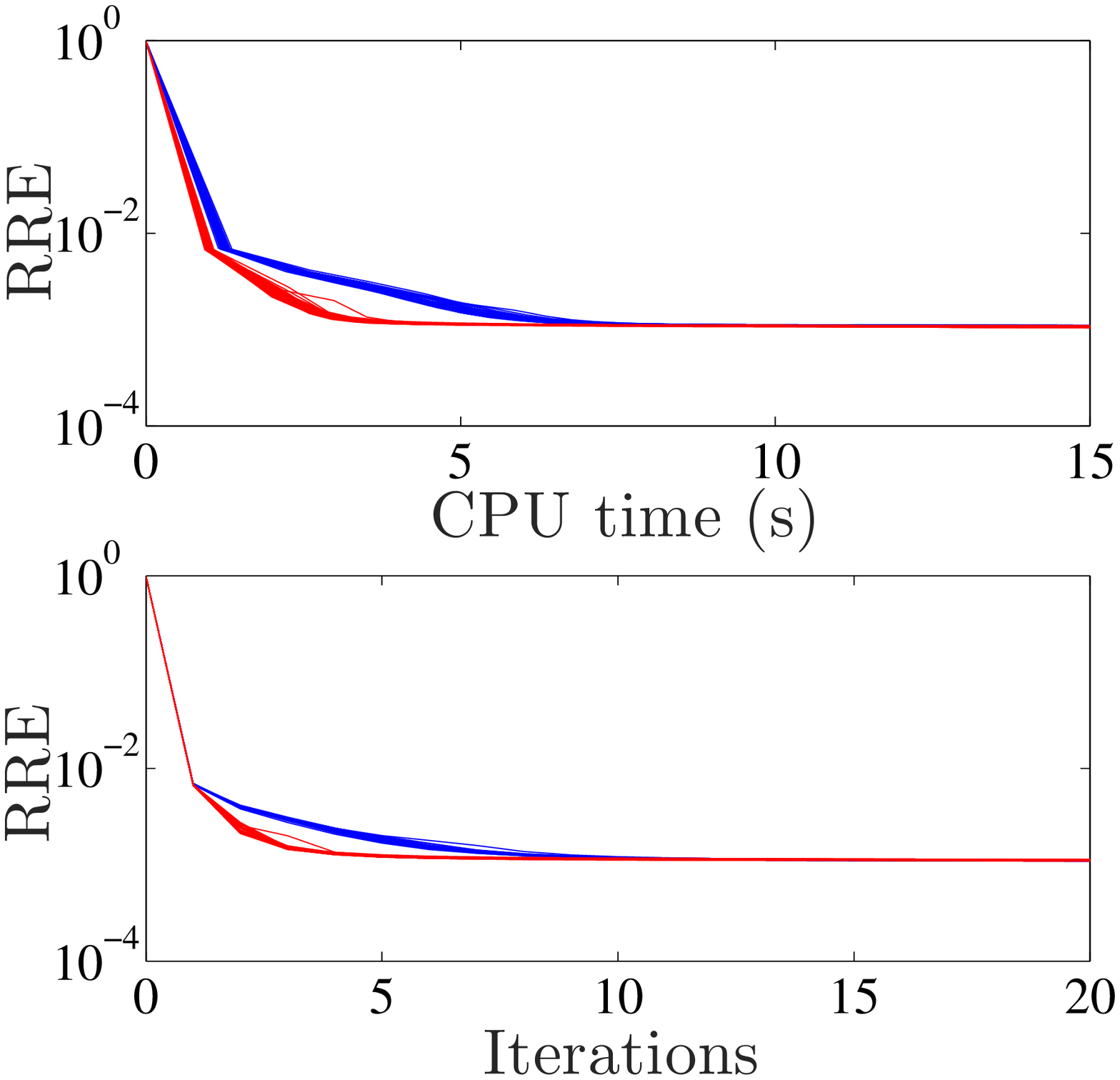}\hfill \includegraphics[width=.265\linewidth]{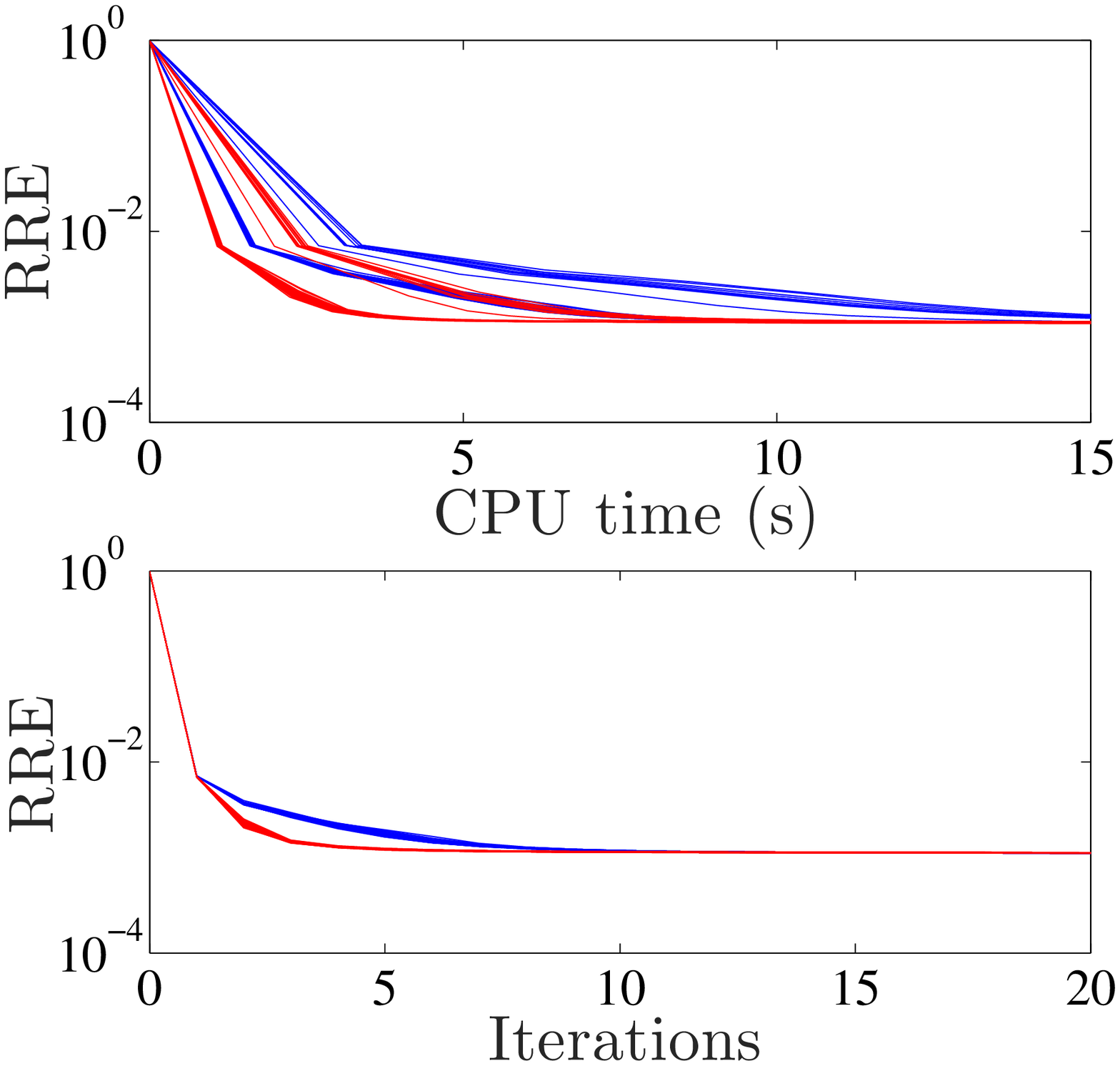}
\caption{Vanilla and compressed NeNMF performance over CPU time (top) and iterations (bottom). Left: $n=500$. Middle: $n=5000$. Right: $n=10000$.}
\label{fig:1}
\end{figure*}

\section{NeNMF with random projections}
\label{sec:first-section}
We now introduce the principles of random projections for NMF. Assuming that the original data matrix $X$ is low-rank, the key idea consists of estimating a smaller matrix---with the same main properties than $X$---whose reduced size allows to fasten the computations. While random projections were initially proposed for singular value decompositions \cite{Halko_2011}, they were more recently applied to NMF in, e.g., \cite{Wang_2010,Tepper_2016,Erichson_2018}. Starting from a target rank $\nu$ (with $p \leq \nu \ll \min(n,m)$), the initial random projection technique consists of drawing scaled\footnote{Denoting $\Omega$ a $ \nu \times n$ random matrix, its scaled version $L$ reads $L = \Omega / \sqrt{ \nu}$.} Gaussian random matrices $L \in \R^{\nu \times n}$ and $R \in \R^{m \times  \nu}$ whose product on the left side of $X$ and $G$, or on the right side of $X$ and $F$, respectively, allows to compress the matrices. The whole strategy is shown in Algorithm~\ref{algo:compressed_nmf}. Please note that as $L$ and $R$ have no sign constraint, the matrices $X_L$, $G_L$, $X_R$, and $F_R$ can get negative entries, so that the update rules in Algorithm~\ref{algo:compressed_nmf} are instances of semi-NMF \cite{Ding_2010}. Lastly, the NMF stopping criterion might be a target approximation error, e.g., 
a reached CPU time.
\begin{algorithm}
\caption{Compressed NMF strategy\label{algo:compressed_nmf}}
\begin{algorithmic}
\REQUIRE initial and compression matrices $G$, $F$, $L$, and $R$.
\STATE Define $ X_L \triangleq L \cdot X$ and $X_R \triangleq X \cdot R$
\REPEAT
\STATE Define $F_R \triangleq F \cdot R  $
\STATE Solve (\ref{eq:NMF_update1}) by resp. replacing $X$ and $F$ by $X_R$ and $F_R$
\STATE Define $G_L \triangleq L \cdot G$
\STATE Solve (\ref{eq:NMF_update2}) by resp. replacing $X$ and $G$ by $X_L$ and $G_L$
\UNTIL{a stopping criterion}
\end{algorithmic}
\end{algorithm}

Actually, the design of randomized compression matrices $L$ and $R$ can be improved, e.g., using randomized power iterations or its stable variant named randomized subspace iteration \cite{Halko_2011}. Applied to NMF, the former was proposed in \cite{Tepper_2016}---under the name of "structured random projection"---while %
the later is summarized in Algorithm~\ref{algo:compression2}. In this configuration, $L$ and $R$ exhibit orthonormal columns and rows, respectively. 
To the best of the authors' knowledge, randomized subspace iteration was never applied to NMF before and we investigate its behaviour in this paper.
%
%
%
%
%

As explained above, the NMF techniques applied to dense matrices $X$ in \cite{Wang_2010,Tepper_2016,Erichson_2018} were using MU, PG, and HALS, respectively. In this paper, we propose to replace them by Nesterov iterations. The main interest of such an investigation consists of seeing whether/when random projections provide some benefits to NMF using an optimal solver.

\section{Experimental validation}\label{sec:third-section}
To investigate the performance of the proposed method, we 
draw random nonnegative matrices $G$ and $F$, with $p=15$ and $n=m$ (with $n\gg p$), such that $X$ is a square low-rank matrix\footnote{We also tested other ranks in some preliminary work without noticing any major differences. 
}. In our tests, we set $n$ to 500, 5000, and 10000, respectively. For each tested value of $n$, we draw 40 different theoretical matrices $G$ and $F$. We also add noise to the observed matrices so that the signal-to-noise ratio is around 30~dB.

\begin{algorithm}
\caption{Randomized subspace iterations for NMF\label{algo:compression2}}
\begin{algorithmic}
\REQUIRE a target rank $\nu$ (with $p \leq \nu \ll \min(n,m)$) and an integer $q$ (e.g., $q=4$)
\STATE Draw Gaussian random matrices $\Omega_L \in \R^{m \times \nu}$ and $\Omega_R \in \R^{\nu \times n}$ 
\STATE Form $ \X_L^{(0)}  \triangleq X \cdot \Omega_L$ and  $ \X_R^{(0)}  \triangleq \Omega_R \cdot X$ 
\item Compute their respective orthonormal bases  $Q_L^{(0)}$ and $Q_R^{(0)}$, by QR decomposition of $\X_L^{(0)}$ and $\X_R^{(0)}$, respectively
\FOR{$k=1$ \TO $q$}
\STATE Define $ \tilde{\X}_L^{(k)}  \triangleq X^T \cdot Q_L^{(k-1)}$ and $ \tilde{\X}_R^{(k)}  \triangleq  Q_R^{(k-1)} \cdot X^T$ 
\STATE Derive their respective orthonormal bases  $\tilde{Q}_L^{(k)} $ and $\tilde{Q}_R^{(k)} $ 
\STATE Compute $ \X_L^{(k)}  \triangleq X \cdot \tilde{Q}_L^{(k)}$ and $ \X_R^{(k)}  \triangleq \tilde{Q}_R^{(k)} \cdot X$ 
\STATE Derive their respective orthonormal bases $Q_L^{(k)} $ and $Q_R^{(k)} $ 
\ENDFOR 
\STATE Derive $L \triangleq \tilde{Q}_L^{(q)}$ and $R \triangleq \tilde{Q}_R^{(q)}$, respectively.
\end{algorithmic}
\end{algorithm}

We only investigate the performance of the NeNMF method without or with the randomized subspace iteration\footnote{In preliminary tests, we compared its performance with MU-NMF, PG-NMF, and HALS-NMF and found it to be much faster. Moreover, we also found the standard random projections \cite{Wang_2010} to provide a lower enhancement than both other compression techniques, which is consistent with \cite{Tepper_2016} for randomized power iteration. Lastly, we found the randomized subspace iteration to slightly outperform the randomized power iteration. Due to space restrictions, we cannot reproduce these results in this paper.}. The target rank of the random matrices is set to $\nu=25$. The tested methods are run during 15~s and, at each NMF iteration, we estimate the relative reconstruction error (RRE), defined as
\begin{equation}
\text{RRE} \triangleq \normF{X - G \cdot F}/ \normF{X},
\end{equation}
over the CPU time. All the methods are run using Matlab R2016a on a laptop with an Intel Core i7-4800MQ Quad Core processor, and 32~GB RAM memory.

Figure~\ref{fig:1} shows the achieved performance by the vanilla NeNMF and its compressed extension, for the different tested values of $n$ and when the maximum number of iterations per NeNMF inner loop is set to $\text{Max}_{\text{iter}} = 500$. In addition to the RRE evolution over the CPU time, we also plot the RRE versus the NMF iterations. The fast RRE decreasing which is visible in the early iterations in each plot 
shows the interest of applying random projections to an already fast NMF technique\footnote{It should be noticed that the compressed NeNMF is sensitive to the value of $\text{Max}_{\text{iter}}$. Indeed, when $\text{Max}_{\text{iter}} = 100$ in \cite{Yahaya_2018}, both the compressed and vanilla NeNMF are faster (in terms of CPU time but not w.r.t. iterations) but the RRE is not always decreasing with the compressed version. This issue might be solved by adaptive restart strategies for example \cite{Odonoghue_2015}.}.
\section{Conclusion}
\label{sec:conclusion}
\MPMP{In this paper, we proposed an NMF method which combines random projections and optimal gradient descent. The proposed method is shown to be (much) faster than vanilla NeNMF. 
In future work, we aim to apply random projections to weighted \cite{Dorffer_2017a} and informed \cite{Dorffer_2018} NMF.}

\paragraph{Acknowledments.} F. Yahaya gratefully acknowledges the R\'egion Hauts-de-France to partly fund his PhD fellowship.

\bibliographystyle{ieeetr}
{\small
    \bibliography{IEEEabrv,biblio.bib}
}

\end{document}